\documentclass[aps,pra,twocolumn,showpacs,superscriptaddress]{revtex4}
\usepackage{bm,amsmath,amssymb,latexsym,mathrsfs,graphicx,enumerate}
\usepackage[mathcal]{euscript}
\usepackage{hyperref}
\usepackage{epsfig}
\usepackage{epstopdf}

\begin{document}

\title{\textbf{Surface Waves on the Interface Between Hyperbolic Material and
Topological Insulator}}
\author{Ekaterina I. Lyashko}
\affiliation{\normalsize \noindent Department of General and Applied Physics,
Moscow Institute of Physics and Technology, Dolgoprudny, Moscow
region, 141700 Russia \\
E-mails: ostroukhova.ei@gmail.com
}
\author{Andrei I. Maimistov}
\affiliation{\normalsize \noindent Department of Solid State Physics
and Nanostructures, National
Nuclear Research University,\\
Moscow Engineering Physics Institute, Moscow, 115409 Russia \\
E-mails: aimaimistov@gmail.com }
 \affiliation{\normalsize \noindent
Department of General Physics,
Moscow Institute of Physics and Technology, Dolgoprudny, Moscow region, 141700 Russia \\
}
\author{Ildar R. Gabitov}
\affiliation{\normalsize \noindent  Department of Mathematics, University of Arizona, \\
Tucson, Arizona, 85721, USA
}
\affiliation{\normalsize \noindent Skolkovo Institute of Science and Technology, Skolkovo Innovation Center, \\ Moscow 143026  Russia \\
E-mails: gabitov@math.arizona.edu
 }

\date{\today}

\begin{abstract}
\noindent Propagation of the surface waves on the interface
 between an uniaxial hyperbolic material and an isotopical topological insulator is studied.
The cases of the anisotropy axes is normal to interface or one is
coplanar to interface are discussed. The dispersion relations are
derived and analyzed. The conditions of the existence of the surface
waves are established.

\end{abstract}

\pacs{42.82.-m, 42.82.Et, 42.79.Gn, 78.67.Pt, 42.25.Bs}


\maketitle

\section{Introduction}

\noindent During last several years considerable attention of
scientific community was attracted to new materials known in
condensed matter physics as topological insulators
(TI)~\cite{Fu:Kane:07,Hasan:Kane:10,Qi:Zhang:11}. Electrons in TI
are mobile only  on the surface of the material and surface states
are protected by time reversal symmetry and conservation law for
number of  particles. Crystals of $\mathrm{Bi_2Te_3}$,
$\mathrm{Bi_2Se_3}$, $\mathrm{Sb_2Te_3}$ doped with  $\mathrm{Fe}$
are representing  examples of such material.  Investigation of
TI~\cite{MingCheChang:09,Reijnders:14}, as well as
study of optical/electrodynamic characteristics of these materials
become a challenging problem of optics. Magneto-electric effect
~\cite{Obukhov:Hehl:05,XiaoLiangQi:08,Essin:09,Tse:10,Moore:15,Morimoto:15},
which manifests itself as a giant Faraday  and Kerr
effect~\cite{Karch:09,Tse:10,FenLiu:14}, is a characteristic feature
of the interaction of electromagnetic field with  TI. Refraction and
electromagnetic wave scattering at the surface of TI are considered
in~\cite{MingCheChang:09,Obukhov:Hehl:05,Ochiai:12,YanyanZhao:14}.
Results of the study of Goos--H\"{a}nchen and Fedorov's shifts on
the surface of TI are presented in the
papers~\cite{MingCheChang:09,Xun:Sun:13}.

It is known that surface waves do not exist on the interface of
topological insulator and conventional dielectric~\cite{Karch:11}.
However, this limitation can be removed by replacing of conventional
dielectric on   other materials. For example  surface wave in the
form of surface plasmon can propagate along the interface of the
topological insulator and metal ~\cite{Karch:11}. Surface waves can
also exist on the interface of topological insulator and
metamaterial with negative index of refraction. These examples
indicate that presence of surface waves can be achieved by use of
material with negative dielectric permittivity $\varepsilon < 0$.
Hyperbolic
materials~\cite{Narimanov:06,Noginov:09,Drachev:13,Kivshar:13,Othman:13}
represent  the broad class of  such
metamaterials~\cite{Elefth:05,Agran:Gart:06,Noginov:12}. They have a
strong uniaxial anisotropy -- principal dielectric permittivities
have opposite signs. The result of this uniaxial anisotropy is a
hyperbolic shape in a $k$-space of the surface  corresponding to a
constant frequency $\omega(\vec{k})=const$ (the iso-frequency
surface). Note that this shape is an ellipsoid for conventional
dielectrics

This paper considers  propagation of the surface wave along the
interface between two homogeneous media: the uniaxial hyperbolic
material and topological insulator. The principal equations
describing this  wave are presented in Sec.~\ref{sec:mequati}. In
accordance with planar symmetry of considered problem,
electromagnetic waves can be studied separately as TE or TM waves
depending on their polarization. However, TE and TM waves are mixing
at the interface  due to nontrivial boundary conditions. In other
words, the wave propagating along the interface is the result of
hybridization of TE and TM waves. The dispersion relation  for
surface wave is derived in Sec. \ref{sec:disp}. In the particular
case, when the constant of magnetoelectric interaction is set to
zero, the dispersion relation transforms to  usual dispersion
relations separately for each type of polarization.

\section{Model of the interface and the principal equations}\label{sec:mequati}

\noindent In the absence of free charges and currents the
macroscopic equations describing the electromagnetic field in TI,
take the following form \cite{Wilczek:87,XiaoLiangQi:08}
\begin{eqnarray}
&&\mathrm{rot}\mathop{\mathbf{H}} - \frac{1}{c}\frac{\partial
\mathbf{D}}{\partial t} =  - \alpha \left( \nabla \theta
\times\mathbf{E} +\frac{1}{c}\frac{\partial
\theta}{\partial t}~\mathbf{B}\right), \nonumber \\
&& \mathrm{rot}\mathop{\mathbf{E}}+\frac{1}{c}\frac{\partial
\mathbf{B}}{\partial t}=0, \label{Axion:macro:field:gene} \\
&& \mathrm{div} \mathop{\mathbf{D}} = \alpha (\nabla \theta \cdot
\mathbf{B}), \quad \mathrm{div} \mathop{\mathbf{B}} =0, \nonumber
\end{eqnarray}
where $\alpha$ is equal to $e^2/\hbar c$ and denotes the coupling
constant of electromagnetic field with the axion field $\theta$
\cite{Wilczek:87}. The product $\alpha\theta$ plays a role of the
magnetoelectrical susceptibility \cite{XiaoLiangQi:08}.

In these equations  vectors $\mathbf{D}=\mathbf{E}+4\pi \mathbf{P}$
and $\mathbf{H}=\mathbf{B}-4\pi \mathbf{M}$ in the absence of
topological effects are equivalent to vectors of electric and
magnetic inductions respectively. They specify the electromagnetic
waves propagation in the conventional dielectric where
electromagnetic wave dynamic is governed by the following system of
equations
\begin{eqnarray}
&& \mathrm{rot}\mathop{\mathbf{E}}=-\frac{1}{c}\frac{\partial
\mathbf{B}}{\partial t},\quad \mathrm{div} \mathop{\mathbf{B}} =0,
\label{macro:field:1}\\
 && \mathrm{rot}\mathop{\mathbf{H}} =
\frac{1}{c}\frac{\partial \mathbf{D}}{\partial t},
 ~~\quad \mathrm{div} \mathop{\mathbf{D}} = 0. \nonumber
\end{eqnarray}

It should be pointed out that if the electric and magnetic
inductions are defined as follows
\begin{equation}
\mathbf{D}_a= \mathbf{D} - \alpha\,\theta\mathbf{B} ,
~~\mathbf{H}_{a}= \mathbf{H} + \alpha\,\theta\mathbf{E},
\label{Axion:macro:inductions}
\end{equation}
then the Maxwell equations (\ref{Axion:macro:field:gene}) are taking
form similar to the system of equations (\ref{macro:field:1})
\begin{eqnarray}
&& \mathrm{rot}\mathop{\mathbf{E}}=-\frac{1}{c}\frac{\partial
\mathbf{B}}{\partial t},~~~\quad \mathrm{div} \mathop{\mathbf{B}}
=0,
\label{Axion:macro:A:Ha:2}\\
 &&   \mathrm{rot}\mathop{\mathbf{H}_a} =
\frac{1}{c}\frac{\partial \mathbf{D}_a}{\partial t},
 ~~\quad \mathrm{div} \mathop{\mathbf{D}_a} = 0. \nonumber
\end{eqnarray}

However, if $\theta$ is constant the Maxwell equations
(\ref{Axion:macro:field:gene}) does not contain the additional
terms. The system of equations (\ref{Axion:macro:field:gene}) (as
well as (\ref{Axion:macro:A:Ha:2})) is completely equivalent to the
conventional Maxwell equations (\ref{macro:field:1}). In this case
the electric and magnetic fields into TI can be derived from the
Maxwell equations (\ref{macro:field:1}). The continuity conditions
on the interface TI-dielectric contain the normal components of the
inductions $\mathbf{D}_a$ and $\mathbf{H}_a$ because they are result
of (\ref{Axion:macro:A:Ha:2}).

The electromagnetic waves in an anisotropic medium are described by
the Maxwell equations (\ref{macro:field:1}), 
where the vector of electric induction $\mathbf{D}$ depends only on
(and non-collinear with)  the $\mathbf{E}$. In the case of
nonmagnetic media the magnetic field and magnetic induction are
equivalent. Fourier transform of the the electric induction vector
in the case of homogeneous uniaxial anisotropic medium has the
following form
\begin{equation}\label{eq:D:anizo:1}
\mathbf{D}=\varepsilon_o\mathbf{E}+
\left(\varepsilon_e-\varepsilon_o\right)(\mathbf{l}\cdot
\mathbf{E})\mathbf{l}.
\end{equation}
Here $\mathbf{l}$ is the unit vector determined by  the optical
axis, $\varepsilon_0=\varepsilon_0(\omega)$ is the permittivity  for
ordinary wave, and $\varepsilon_e= \varepsilon_e(\omega)$ is the
permittivity for extraordinary wave. If the sings of the principal
permittivities are opposite, then the iso-frequency surface is a
hyperboloid: the single sheeted hyperboloid under condition
$\varepsilon_e>0$, $\varepsilon_o<0$ and two sheeted  hyperboloid
under condition $\varepsilon_e<0$, $\varepsilon_o>0$. Such materials
are referred to the hyperbolic metamaterial
\cite{Noginov:09,Drachev:13,Kivshar:13,Othman:13}.

Both equations for the TI and the hyperbolic material can be
separately solved using standard methods. Obtained solutions must be
matched on the interface. The matching procedure is determined by
continuity condition for tangent components of the electric fields
and the normal components of the inductions \cite{Born:Wolf}. It is
important to remember that these continuity conditions follows from
the system of equations (\ref{Axion:macro:A:Ha:2}).

\begin{figure}[h]
    \centering
   \includegraphics[scale=0.5]{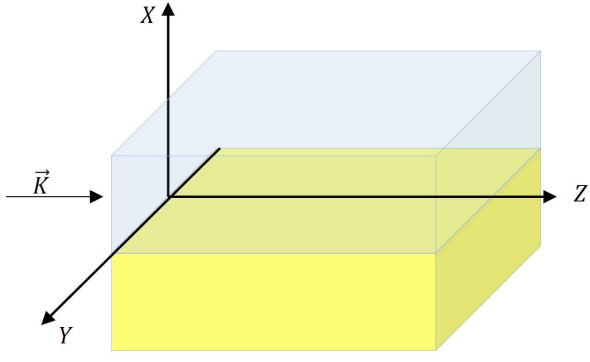}
    \caption{A schematic illustration of the interface between TI slab on hyperbolic substrate.}
    \label{fig:1}
\end{figure}

Let us assume that coordinate axis $X$ is directed along the vector
$\mathbf{n}$ normal to the interface and axes $Y$ and $Z$ are
directed along two orthogonal  vectors $\mathbf{t}_y$ and
$\mathbf{t}_z$ which are tangential to this interface. Let the
region $x<0$ is filled  by the hyperbolic material and  the
topological insulator fills the region $x>0$, see Fig. \ref{fig:1} .
Under these conditions the continuity conditions take the form
\begin{eqnarray}
  && (\mathbf{D}^{(1)}-\mathbf{D}^{(2)})\cdot \mathbf{n} = -\alpha\theta
  \mathbf{B}^{(1)}\cdot \mathbf{n},  \nonumber\\
  && (\mathbf{H}^{(1)}-\mathbf{H}^{(2)})\cdot \mathbf{t}_{z,y} = \alpha \theta
  \mathbf{E}^{(1)}\cdot \mathbf{t}_{z,y},   \label{eq:DandT:sw:contin}\\
  && (\mathbf{B}^{(1)}-\mathbf{B}^{(2)})\cdot \mathbf{n} =0, \quad
  (\mathbf{E}^{(1)}-\mathbf{E}^{(2)})\cdot \mathbf{t}_{z,y} =0.\nonumber
\end{eqnarray}
Here the upper index 1 marks the region $x<0$, and upper index 2
marks the region $x>0$. For TI $\theta$  is known to be $\theta=
2n+1$, where $n \in \mathbb{Z}$ \cite{XiaoLiangQi:08} and $\theta=0$
is for dielectric.

Surface wave propagating along $Z$ does not depend on variable $y$
due to translation symmetry (see Fig~\ref{fig:1}). In this case the
system of Maxwell equations  reads as
\begin{eqnarray}
  && ik_0H_x=-\frac{\partial E_y}{\partial z}, \qquad  ik_0H_z = \frac{\partial E_y}{\partial x}, \nonumber \\
  && \qquad -ik_0D_y = \frac{\partial H_x}{\partial z} -\frac{\partial H_z}{\partial x},
    \label{Axion:TE:field:1}\\
  && ik_0D_x= \frac{\partial H_y}{\partial z}, \qquad  ik_0D_z = -\frac{\partial H_y}{\partial x}, \nonumber\\
  && \qquad ~~ ik_0H_y = \frac{\partial E_x}{\partial z} -\frac{\partial E_z}{\partial x},
    \label{Axion:TM:field:1}
\end{eqnarray}

In the hyperbolic medium at $x<0$ the relation between components of
the induction vector and components of the electric field vector
depends on the orientation of the optical axis $\mathbf{l}$. Two
particular cases for this orientation will be considered further.

\section{Dispersion relations. Electric and magnetic field distributions}\label{sec:disp}

\noindent Direction of the optical axis in hyperbolic medium is
determined by the method of its fabrication.  In the most cases such
materials are structured as alternating flat sheets of metal and
dielectric~\cite{Jacob:06,Kim:12,Othman:13,Kildishev:14} and as 3D
array of  metallic rods in dielectric
matrix~\cite{Narimanov:06,Noginov:09,Xiang:15}. According to theses
the method of the hyperbolic medium fabrication optical axis are
orthogonal or tangential to the media interface. Both cases cases
will be analyzed 
.

\subsection{Optical axis is tangential  to the interface.
}\label{subsec:disp:y}

\noindent In the domain occupied by anisotropic medium, when optical
axis is tangential to the interface  and orthogonal to propagation
direction $\mathbf{l}=\mathbf{t}_y $, components of the induction
vector take the following form
$$
D_x=\varepsilon_oE_x, \quad D_y=\varepsilon_eE_y, \quad
D_z=\varepsilon_oE_z.
$$

The first group of equations from the (\ref{Axion:TE:field:1}), that
can be named as equations of the TE-type, takes the following
form
\begin{eqnarray}
  && \frac{\partial^2 E_y}{\partial z^2}+ \frac{\partial^2 E_y}{\partial x^2}+ k_0^2 \varepsilon_e E_y =0,
   \label{Axion:TE:field:2}\\
  &&  H_x=\frac{i}{k_0}\frac{\partial E_y}{\partial z}, \quad  H_z = -\frac{i}{k_0}\frac{\partial E_y}{\partial x},
  \nonumber
\end{eqnarray}
The second group of  equations named as TM-type equations can be  rewritten as
\begin{eqnarray}
  && \frac{\partial^2 H_y}{\partial z^2}+ \frac{\partial^2 H_y}{\partial x^2}+ k_0^2 \varepsilon_o H_y =0,
   \label{Axion:TM:field:2}\\
  &&  E_x=-\frac{i}{k_0\varepsilon_o}\frac{\partial H_y}{\partial z}, \quad  E_z = \frac{i}{k_0\varepsilon_o}\frac{\partial H_y}{\partial x},
  \nonumber
\end{eqnarray}

The surface wave is characterized by the boundary conditions
$\mathbf{E} \to 0$ and $\mathbf{H} \to 0 $ at $x \to \mp\infty$. The
solutions of the TE-type equations obeying to these conditions when $x \to -\infty$ is
possible if $ p_1^2=\beta^2-k_0^2\varepsilon_e >0 $. The electric
and magnetic fields read
\begin{eqnarray}
  H_x^{(1)}(x,z) &=& -\frac{\beta}{k_0}Ae^{p_1x+i\beta z},  \nonumber\\
  H_z^{(1)}(x,z) &=& -\frac{ip_1}{k_0}Ae^{p_1x+i\beta z},\label{Axion:TE:field:3} \\
  E_y^{(1)}(x,z) &=& Ae^{p_1x+i\beta z}.\nonumber
\end{eqnarray}

The solutions of the TM-type equations obeying to the boundary
conditions is possible if  $ p_2^2=\beta^2-k_0^2\varepsilon_o
>0$. The electric and magnetic fields are
\begin{eqnarray}
  E_x^{(1)}(x,z) &=& \frac{\beta}{k_0\varepsilon_o}Be^{p_2x+i\beta z},  \nonumber\\
  E_z^{(1)}(x,z) &=& \frac{ip_2}{k_0\varepsilon_o}Be^{p_2x+i\beta z},\label{Axion:TM:field:3} \\
  H_y^{(1)}(x,z) &=& Be^{p_2x+i\beta z}.\nonumber
\end{eqnarray}

In the domain occupied by TI, which is assumed to be an
isotropic medium, the components of the induction vector
take the following form
$$
D_x=\varepsilon_2E_x, \quad D_y=\varepsilon_2E_y, \quad
D_z=\varepsilon_2E_z.
$$
The wave equations in this case have the form similar to  equations
(\ref{Axion:TE:field:2}) and (\ref{Axion:TM:field:2}), where
$\varepsilon_o$ and $\varepsilon_e$ are replaced by
$\varepsilon_2$.

Implementation of the boundary conditions $\mathbf{E} \to 0$ and
$\mathbf{H} \to 0 $ at $x \to +\infty$ for solutions of the TE- and
TM-type of equations is possible if $ q^2=\beta^2-k_0^2\varepsilon_2
>0$. The electric and magnetic fields have the following form
\begin{eqnarray}
  H_x^{(2)}(x,z) &=& -\frac{\beta}{k_0}Ce^{-q x+i\beta z},  \nonumber\\
  H_z^{(2)}(x,z) &=& \frac{iq}{k_0}Ce^{-q x+i\beta z},\label{Axion:TE:field:4} \\
  E_y^{(2)}(x,z) &=& Ce^{-q x+i\beta z}.\nonumber
\end{eqnarray}
\begin{eqnarray}
  E_x^{(2)}(x,z) &=& \frac{\beta}{k_0\varepsilon_2}Fe^{-q x+i\beta z},  \nonumber\\
  E_z^{(2)}(x,z) &=& -\frac{iq}{k_0\varepsilon_2}Fe^{-q x+i\beta z},\label{Axion:TM:field:4} \\
  H_y^{(2)}(x,z) &=& Fe^{-q x+i\beta z}.\nonumber
\end{eqnarray}

The amplitudes $A$, $B$, $C$ and $F$ can be found from the
continuity conditions on the interface (\ref{eq:DandT:sw:contin}).
Substitution of
(\ref{Axion:TE:field:3})--(\ref{Axion:TM:field:4}) into
(\ref{eq:DandT:sw:contin}) results in following system of the
algebraic equations
\begin{eqnarray}
  && qC+p_1A=\kappa\frac{p_2}{\varepsilon_o}B,\qquad B-F=-\kappa A,  \nonumber \\
  && \frac{p_2}{\varepsilon_o}B +\frac{q}{\varepsilon_2}F=0,\qquad\quad  A-C=0,\nonumber
\end{eqnarray}
where $\kappa = \alpha\theta$. Nontrivial solutions of these equations
exist if  determinant of this system of equations is zero.
This requirement results in the dispersion relation
$$
(q+p_1)\left(1+ \frac{p_2\varepsilon_2}{q\varepsilon_o} \right) +
\kappa^2\frac{p_2}{\varepsilon_o}=0,
$$
that can be rewritten as
\begin{equation}\label{Axion:dispers:TIHM:1}
    (q+p_1)\left(\frac{q}{\varepsilon_2}+ \frac{p_2}{\varepsilon_o}
\right) + \kappa^2\frac{p_2q}{\varepsilon_o\varepsilon_2}=0.
\end{equation}

According to definitions the conditions $q>0$, $p_1>0$ and $p_2>0$
are held. If $\varepsilon_o> 0$ and $\varepsilon_2> 0$ the equation
(\ref{Axion:dispers:TIHM:1}) has no solution, as it is sum of the
positive terms. However, if $\varepsilon_o < 0$ and $\varepsilon_2>
0$ the equation (\ref{Axion:dispers:TIHM:1}) takes the form of
\begin{equation}\label{Axion:dispers:TIHM:2}
    (q+p_1)\left(\frac{q}{\varepsilon_2}- \frac{p_2}{|\varepsilon_o| }
\right)- \kappa^2\frac{p_2q}{|\varepsilon_o|\varepsilon_2}=0
\end{equation}
and this equation is soluble.

Thus, in the case of hyperbolic material with $\varepsilon_o < 0$
and $\varepsilon_e > 0$, surface wave can  exist, provided that
the optical axis is directed normally to the direction of propagation and
lies in interface.

\subsection{Optical axis is normal to interface.
}\label{subsec:disp:x}

\noindent The components of the electric induction vector take the
following form
$$
D_x=\varepsilon_eE_x, \quad D_y=\varepsilon_oE_y, \quad
D_z=\varepsilon_o E_z.
$$
Having the relation $D_y=\varepsilon_oE_y$  the TE-type
equations take the form
\begin{eqnarray}
  && \frac{\partial^2 E_y}{\partial z^2}+ \frac{\partial^2 E_y}{\partial x^2}+ k_0^2 \varepsilon_o E_y =0,
   \label{Axion:TE:field:b2}\\
  &&  H_x=\frac{i}{k_0}\frac{\partial E_y}{\partial z}, \quad  H_z = -\frac{i}{k_0}\frac{\partial E_y}{\partial x}.
  \nonumber
\end{eqnarray}
Solutions satisfying to the surface wave conditions  read as follows
\begin{eqnarray}
  H_x^{(1)}(x,z) &=& -\frac{\beta}{k_0}Ae^{p_1x+i\beta z},  \nonumber\\
  H_z^{(1)}(x,z) &=& -\frac{ip_1}{k_0}Ae^{p_1x+i\beta z},\label{Axion:TE:field:b3} \\
  E_y^{(1)}(x,z) &=& Ae^{p_1x+i\beta z},\nonumber
\end{eqnarray}
where $ p_1^2=\beta^2- k_0^2\varepsilon_o
>0 $ and $ p_1 > 0$.

Surface wave solutions of TM-type are
\begin{eqnarray}
  &&\frac{1}{\varepsilon_e} \frac{\partial^2 H_y}{\partial z^2}+ \frac{1}{\varepsilon_o}\frac{\partial^2 H_y}{\partial x^2}+ k_0^2 H_y =0,
   \label{Axion:TM:field:c2}\\
  && E_x=-\frac{i}{k_0\varepsilon_e}\frac{\partial H_y}{\partial z},
\quad  E_z = \frac{i}{k_0\varepsilon_o}\frac{\partial H_y}{\partial
x},
  \nonumber
\end{eqnarray}
have s similar form
\begin{eqnarray}
  E_x^{(1)}(x,z) &=& \frac{\beta}{k_0\varepsilon_e}Be^{p_2x+i\beta z},  \nonumber\\
  E_z^{(1)}(x,z) &=& \frac{ip_2}{k_0\varepsilon_o}Be^{p_2x+i\beta z},\label{Axion:TM:field:b4} \\
  H_y^{(1)}(x,z) &=& Be^{p_2x+i\beta z}.\nonumber
\end{eqnarray}
where $p_2^2
=(\varepsilon_o/\varepsilon_e)(\beta^2-k_0^2\varepsilon_e)
>0$ and
 $p_2>0$.

The solutions of the Maxwell equations in the region filled with  TI
($x>0$) were found in Sec.\ref{subsec:disp:y},  see (\ref{Axion:TE:field:4}) and
(\ref{Axion:TM:field:4}),  and can be used
here. The continuity conditions taking into account expressions
(\ref{Axion:TE:field:4}), (\ref{Axion:TM:field:4}),
(\ref{Axion:TE:field:b3}) and  (\ref{Axion:TM:field:b4})
 lead to the dispersion relation for the case under consideration
$$
(q+p_1)\left(1+ \frac{p_2\varepsilon_2}{q\varepsilon_o} \right) +
\kappa^2\frac{p_2}{\varepsilon_o}=0.
$$
Note, if $\varepsilon_o >0$, the equations has no solution. On
the other hand in case  of  hyperbolic material with
$\varepsilon_o <0$ and $\varepsilon_e >0$,
this dispersion relation admits solution if
$$
p_1^2=\beta^2 +k_0^2|\varepsilon_o| >0, \quad
\frac{|\varepsilon_o|}{\varepsilon_e}(k_0^2\varepsilon_e-\beta^2)
>0.
$$
It follows that surface wave exists under condition
$$\beta^2/k_0^2 < \varepsilon_e.$$

\section{Analysis of the dispersion relations }

\noindent In the following analysis the dimensionless parameter
$n_{eff} = \beta/k_0$ will be used. The inhomogeneous material under
consideration can be considered as the uniform one, which is
characterized by the effective index $n_{eff}$. It is standard
practice to treat the waveguide or fiber systems \cite{Hunsper:84}.
In the case of a transparent medium $n_{eff}$ defines the phase
velocity of the guided wave according to expression $v_{ph}=c/n_{eff}$.

\subsection{Optical axis is tangential to the interface. $\mathbf{l}=\mathbf{t}_y $}

\noindent  The dispersion relation (\ref{Axion:dispers:TIHM:2})
obtained earlier will take a form:
\begin{eqnarray}
&&\label{eq:DispTy} \left(1+\sqrt{\frac{n_{eff}^2-
\varepsilon_e}{n_{eff}^2- \varepsilon_2}} \right) \left(1-
\frac{\varepsilon_2}{|\varepsilon_o|}\sqrt{\frac{n_{eff}^2+
|\varepsilon_o|}{n_{eff}^2
- \varepsilon_2}} \right) = \nonumber\\
&& \qquad \qquad
=\frac{\kappa^2}{|\varepsilon_o|}\sqrt{\frac{n_{eff}^2+
|\varepsilon_o|}{n_{eff}^2- \varepsilon_2}}.
\end{eqnarray}
This relation connects effective index $n_{eff}$ and radiation
frequency $\omega$ implicitly. We suppose that radiation frequency
is far away from all resonance frequencies specific to the
insulator, thus $\varepsilon_2$ is a constant. But we take into
account dispersion of the hyperbolic medium parameters,
$\varepsilon_e(\omega)$ and $\varepsilon_o(\omega)$.

The simplest for realization type of hyperbolic medium is material
formed from alternating conductive and dielectric layers of
sub-wavelength width. In effective medium approach dielectric
constants of such medium can be presented as
\begin{eqnarray}
\label{eq:EffectMedium}
\varepsilon_o = f_m\varepsilon_m + (1-f_m)\varepsilon_d, \\
\nonumber \varepsilon_e = \frac{\varepsilon_d
\varepsilon_m}{f_m\varepsilon_d
    + (1-f_m)\varepsilon_m},
\end{eqnarray}
where $f_m$ is a volume fraction of metal, the filling factor.

\begin{figure}[h!]
    \centering
    \includegraphics[scale=0.6]{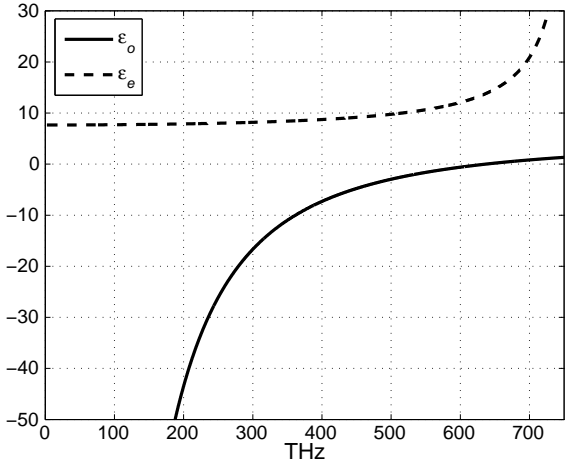}
    \caption{Dependencies of dielectric permittivities of hyperbolic medium on frequency.}
    \label{fig2}
\end{figure}

Usually resonance frequencies of dielectric are lying in the
ultraviolet region. So considering infrared and optical frequencies
of radiation in hyperbolic material one could take into account the
frequency dispersion only of metal layers:
\begin{equation}
\nonumber \varepsilon_d = \mathrm{const}, \quad \varepsilon_m =
\varepsilon_{\infty} -\frac{\omega_p^2}{\omega^2+i\gamma\omega},
\end{equation}
where the Drude-Lorentz model was used. $\omega_p$ is plasma
frequency, $\gamma$ is collision frequency, $\varepsilon_{\infty}$
is permittivity at height frequencies. By constitution of these
relations to the (\ref{eq:EffectMedium}) one can receive the
frequency dependencies  $\varepsilon_e(\omega)$ and
$\varepsilon_o(\omega)$. They are presented in Fig \ref{fig2}.
Parameters were chosen as follows: $\omega_p^2 = 1.38\cdot10^{16}$
rad/c, $\varepsilon_{\infty} = 5$ (Ag), $\gamma = 0$ rad/c,
$\varepsilon_d = 4.6$ (SiO$_2$), $f_m = 0.4$. Metamaterial described
by permittivities (\ref{eq:EffectMedium}) is hyperbolic medium with
$\varepsilon_o < 0$ and $\varepsilon_e > 0$ for chosen frequency
range up to approximately 620 THz.

If $\kappa = 0$ it is possible to consider two types of surface wave
independently. Wave of TE type is described by first multiplier in
the left side of equation (\ref{eq:DispTy}):
$$
1+\sqrt{\frac{n_{eff}^2- \varepsilon_e(\omega)}{n_{eff}^2-
\varepsilon_2}} = 0.
$$
This equation has no real solutions. Thus, surface wave of TE type
is not propagate in the case of linear electrodynamics.
\begin{figure}[h!]
    \centering
    \includegraphics[scale=0.6]{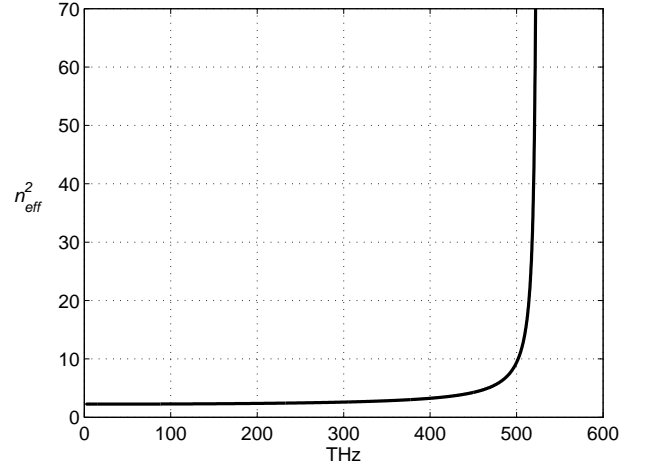}
    \caption{Dependence of $n_{eff}^2$ on frequency at $\kappa = 0$, $\mathbf{l}=\mathbf{t_y}$.}
    \label{fig3}
\end{figure}
Second multiplier in the left side of equation (\ref{eq:DispTy})
describe TM wave:
$$
1-
\frac{\varepsilon_2}{|\varepsilon_o(\omega)|}\sqrt{\frac{n_{eff}^2+
|\varepsilon_o(\omega)|}{n_{eff}^2- \varepsilon_2}} = 0.
$$
This equation can be solved if $|\varepsilon_o| > \varepsilon_2$.
That is because an expression under the square root symbol is always
more than one. Solution of this equation (or of eq.(\ref{eq:DispTy})
at $\kappa = 0$) is presented at Fig. \ref{fig3}. The value of
topological insulator permittivity is $\varepsilon_2 = 2.25$. So
condition $|\varepsilon_o| > \varepsilon_2$ is hold everywhere in
the considering frequency interval. At low frequencies $n_{eff}^2$
tends to $\varepsilon_2$. Also critical value of frequency,
$\omega_c$, at which $n_{eff}^2 \to \infty$ exists. Critical value
$\omega_c$ can be defined from equation $\varepsilon_o(\omega_c) =
-\varepsilon_2$.

Let us consider a topologically nontrivial insulator, i.e. $\kappa
=(2n+1)(e^2/\hbar c) = (2n+1)\alpha$, $n=0,1,..,$. As follows
from the equation (\ref{eq:DispTy}) a surface wave of hybrid
polarization propagates on the media interfaces. A "topological"
term in the right side of this equation connects wave polarizations
of TE and TM type to one hybrid polarization. This implies
additional condition on possible $n_{eff}^2$ values: $n_{eff}^2 >
\varepsilon_e$ besides inequality $n_{eff}^2 > \varepsilon_2$.

Critical frequency value is a function $\omega(\kappa)$ in
considering situation. The definition of $\omega(\kappa)$ could be
obtained from relation (\ref{eq:DispTy}) in case $n_{eff}^2 \to
\infty$:
\begin{equation}
\label{eq:OmegaC} \omega_c^2 =
\frac{f_m\omega_p^2}{f_m\varepsilon_{\infty}+(1-f_m)\varepsilon_d+\varepsilon_2+\kappa^2/2}.
\end{equation}
From equation (\ref{eq:OmegaC}) follows that value of $\omega_c$
shifts to the lower frequencies with $\kappa$. It is interesting to
note, that by substitution $f_m=1$ and $\kappa=0$ the definition
(\ref{eq:OmegaC}) reduces to standard plasmon-polariton critical
frequency value: $\omega_c^2 =
\omega_p^2/(\varepsilon_{\infty}+\varepsilon_2)$.

\begin{figure}[h!]
    \begin{minipage}[ht]{.9\linewidth}
        \centering
        \includegraphics[width=\linewidth]{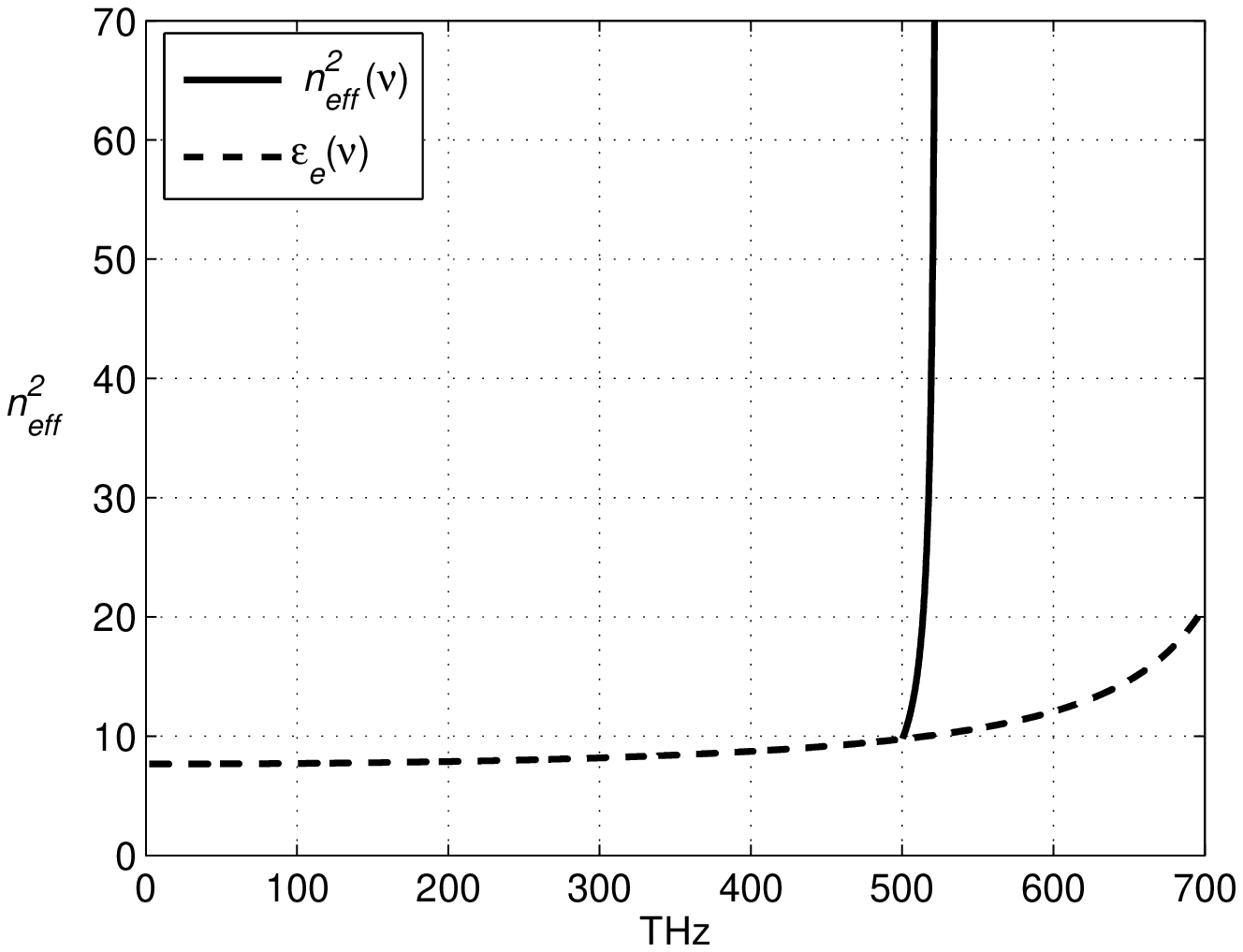}
        \\(\textit{a}) $\kappa = (2\times 10 + 1)\alpha$
    \end{minipage}
    \vfill
    \begin{minipage}[ht]{.9\linewidth}
        \centering
        \includegraphics[width=\linewidth]{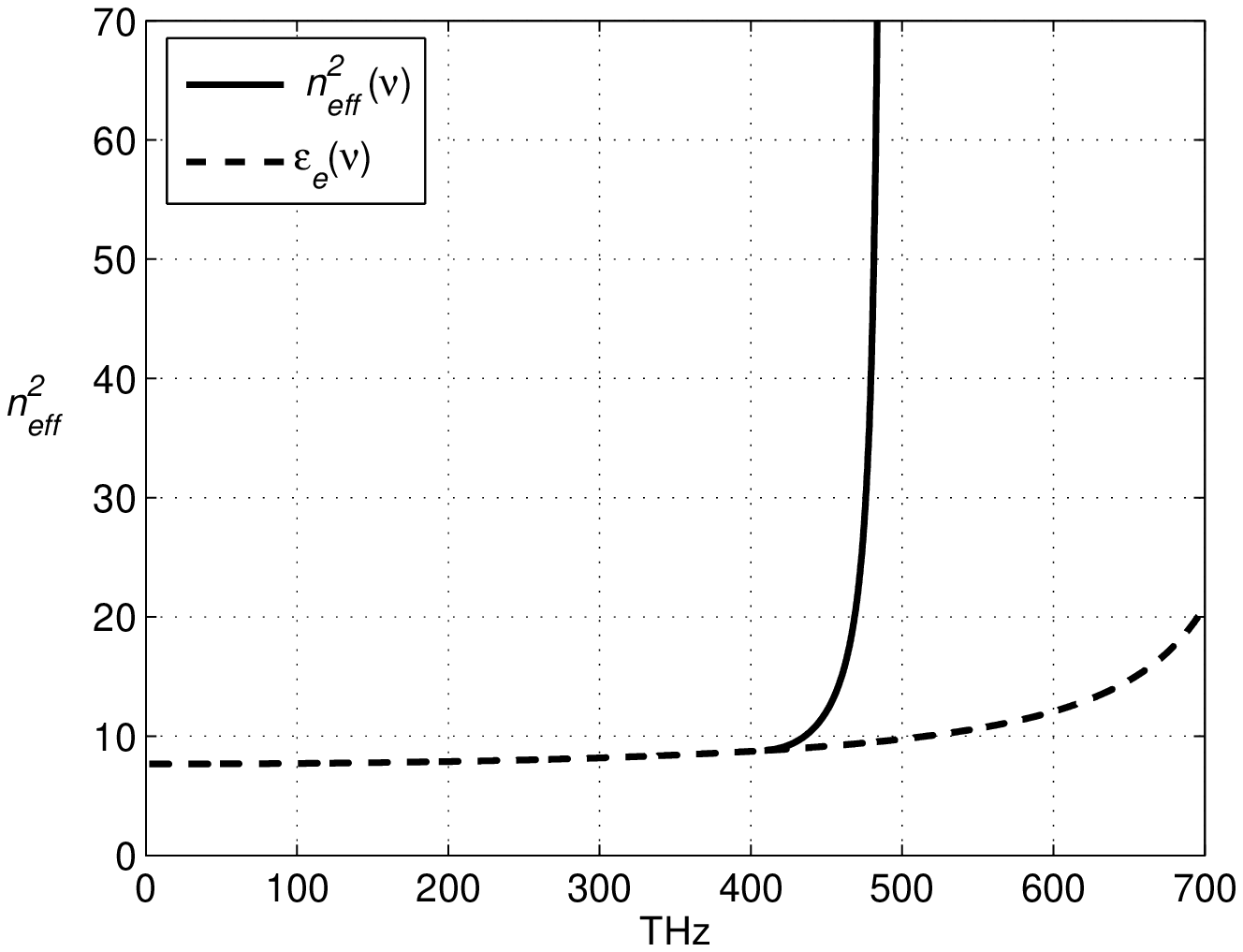}
        \\(\textit{b}) $\kappa = (2\times 100 + 1)\alpha$
    \end{minipage}
    \caption{Dependence of $n_{eff}^2$ on frequency for
    interface with topologically nontrivial insulator, $\varepsilon_e(\nu) > \varepsilon_2$,
     $\mathbf{l}=\mathbf{t_y}$.}
    \label{fig4}
\end{figure}

Solutions of dispersion equation (\ref{eq:DispTy}) at different
values of $\kappa$ are presented in Fig. \ref{fig4}.
In situation presented in Fig. \ref{fig4} an inequality
$\varepsilon_e(\nu) > \varepsilon_2$ is hold for all frequencies
($\varepsilon_2 = 2.25$). Thus, additional condition, $n_{eff}^2 >
\varepsilon_e(\nu)$, leads to the decrease of frequency interval of
surface wave existence. At $\kappa>0$ only narrow enough frequency
domain where $n_{eff}$ is defined exists. From pictures presented in
Fig. \ref{fig4} one can notice that $\omega_c$ (or $\nu_c$ for THz
instead of radians per second) shifts to lower frequencies.

So connection of TM and TE waves at topological insulator/hyperbolic
material interface lead to the rigid restriction on the frequency
domain of surface wave existence at $\varepsilon_e > \varepsilon_2$.
If $\varepsilon_e < \varepsilon_2$ for all considering frequencies,
the surface wave will exist in interval $(0, \omega_c)$ as shown in
Fig. \ref{fig3}. This situation is similar with the case of
topological insulator/metal interface considered in \cite{Karch:11}.
In these cases an existence of nonzero parameter $\kappa$ will lead
only to $\omega_c(\kappa)$ shift to lower frequencies.

\subsection{Optical axis is tangential to the interface, $\mathbf{l}=\mathbf{t}_z $}

\noindent When optical axis is aligned with $Z$ axis the wave
propagation parameters used earlier have the following definitions:
\begin{eqnarray}
  && p_1^2 = k_0^2(n_{eff}^2 - \varepsilon_o),\quad  p_2^2 = k_0^2
\frac{\varepsilon_e}{\varepsilon_o}(n_{eff}^2 - \varepsilon_o),\nonumber \\
  && q^2 = k_0^2(n_{eff}^2 - \varepsilon_2).\nonumber
\end{eqnarray}
All presented parameters must be real and positive.

Using the same technique as presented in previous section a
dispersion relation for situation $\mathbf{l} = \mathbf{t}_z$ can be
achieved. It has a form:
\begin{eqnarray}
\label{eq:DispTz} &&
\left(1+\sqrt{\frac{n_{eff}^2-\varepsilon_o}{n_{eff}^2
-\varepsilon_2}}\right) \left(1+
\frac{\varepsilon_2}{\varepsilon_e}\sqrt{\frac{\varepsilon_e}{\varepsilon_o}\frac{n_{eff}^2
- \varepsilon_o}{n_{eff}^2- \varepsilon_2}} \right) = \nonumber\\
&& \qquad \qquad =
-\frac{\kappa^2}{\varepsilon_e}\sqrt{\frac{\varepsilon_e}{\varepsilon_o}\frac{n_{eff}^2-
\varepsilon_o}{n_{eff}^2- \varepsilon_2}}.
\end{eqnarray}
When $\kappa = 0$ relation (\ref{eq:DispTz}) splits into two
equations describing TE surface wave:
$$
1+\sqrt{\frac{n_{eff}^2-\varepsilon_o}{n_{eff}^2-\varepsilon_2}}= 0,
$$
that has no real solutions, and TM wave:
$$
1+
\frac{\varepsilon_2}{\varepsilon_e}\sqrt{\frac{\varepsilon_e}{\varepsilon_o}\frac{n_{eff}^2-
\varepsilon_o}{n_{eff}^2- \varepsilon_2}} = 0.
$$
The second equation have solutions in case of hyperbolic material
with permittivities $\varepsilon_e < 0$, $\varepsilon_o > 0$. In
case of $\varepsilon_e >0$ both summands are positive and the sum
can't be equal to zero. Taking into account definitions of $p_2$ and
$q$ effective refractive index must satisfy conditions:
$\varepsilon_2 < n_{eff}^2 < \varepsilon_o$. But when a
topologically nontrivial insulator is under consideration, one must
take into account additional condition, comes from the TE wave
description, $n_{eff}^2 > \varepsilon_o$. This inequality is in the
contradiction with previous one. Thus, the relation
(\ref{eq:DispTz}) at $\kappa > 0$ has no real solutions. The surface
wave is impossible on boundary between topological insulator and
hyperbolic media with optical axis aligned with propagation
direction.

\subsection{Optical axis is normal to the interface}

\noindent When optical axis of hyperbolic medium is aligned with
normal to the interface the dispersion relation has solutions at
$\varepsilon_o < 0$, $\varepsilon_e > 0$ and has a form:
\begin{eqnarray}
\label{eq:DispN}
&& \left(1+\sqrt{\frac{n_{eff}^2+
|\varepsilon_o|}{n_{eff}^2-\varepsilon_2}} \right) \left(1-
\frac{\varepsilon_2}{|\varepsilon_o|}\sqrt{\frac{|\varepsilon_o|}{\varepsilon_e}
\frac{\varepsilon_e-n_{eff}^2}{n_{eff}^2 - \varepsilon_2}} \right) = \nonumber\\
&& \qquad \qquad
=\frac{\kappa^2}{|\varepsilon_o|}\sqrt{\frac{|\varepsilon_o|}{\varepsilon_e}\frac{\varepsilon_e
- n_{eff}^2}{n_{eff}^2- \varepsilon_2}},
\end{eqnarray}
where definitions of parameters $p_1, p_2, q$ were used:
\begin{eqnarray}
\nonumber p_1^2 = k_0^2(n_{eff}^2 + |\varepsilon_o|), \\ \nonumber
p_2^2 = k_0^2 \frac{|\varepsilon_o|}{\varepsilon_e}(\varepsilon_e -
n_{eff}^2), \\
\nonumber q^2 = k_0^2(n_{eff}^2 - \varepsilon_2).
\end{eqnarray}

To satisfy conditions $p_1^2 > 0, p_2^2 > 0$ and $q^2 > 0$ a value
of effective refractive index must lie in the interval
$\varepsilon_2 < n_{eff}^2 < \varepsilon_e$. In contrast to previous
cases an additional condition from the TE wave description,
$n_{eff}^2 + |\varepsilon_o| > 0$, is always held. Thus a case of
hyperbolic material/topological insulator boundary has no additional
restrictions in comparison with case of hyperbolic
material/dielectric boundary.

\begin{figure}[h!]
    \begin{minipage}[ht]{.9\linewidth}
        \centering
        \includegraphics[width=\linewidth]{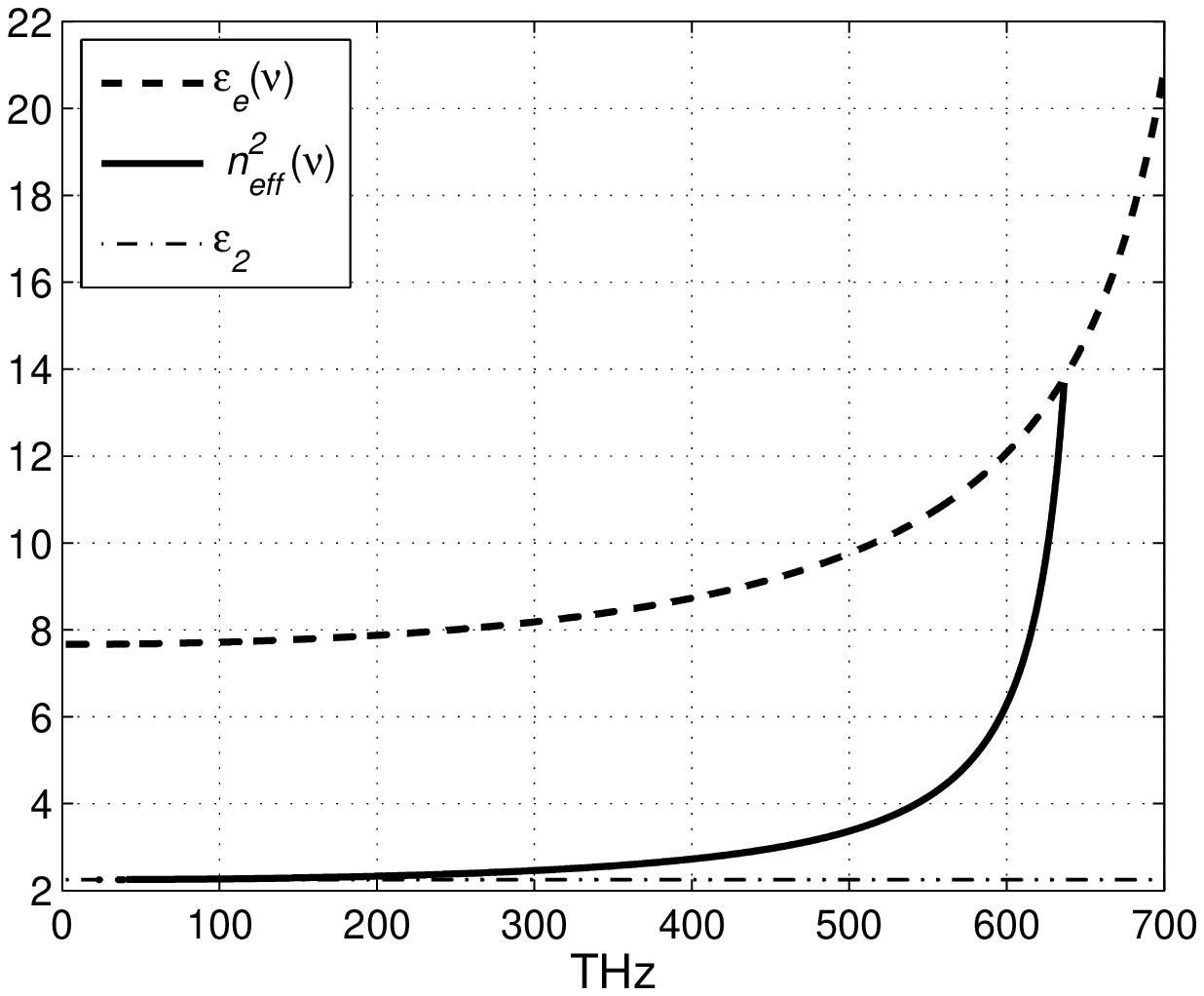}
        \\(\textit{a}) $\kappa = 0$
    \end{minipage}
    \vfill
    \begin{minipage}[ht]{.9\linewidth}
        \centering
        \includegraphics[width=\linewidth]{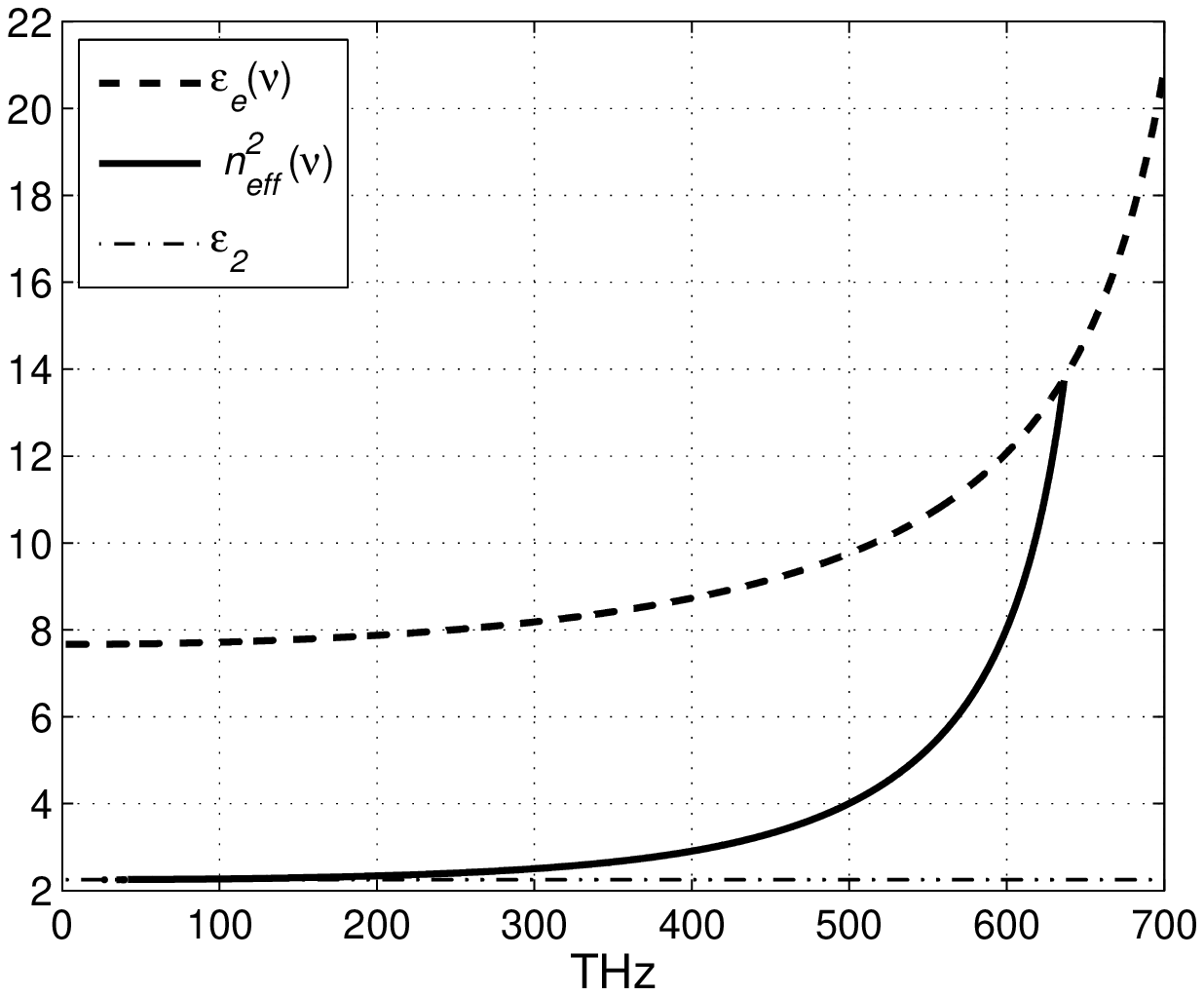}
        \\(\textit{b}) $\kappa = (2\times 100 + 1)\alpha$
    \end{minipage}
    \caption{Dependence of $n_{eff}^2$ on frequency for interface with
    topologically nontrivial insulator, $\mathbf{l} = \mathbf{n}$.}
    \label{fig5}
\end{figure}

\begin{figure}
    \begin{minipage}[ht]{.9\linewidth}
        \centering
        \includegraphics[width=\linewidth]{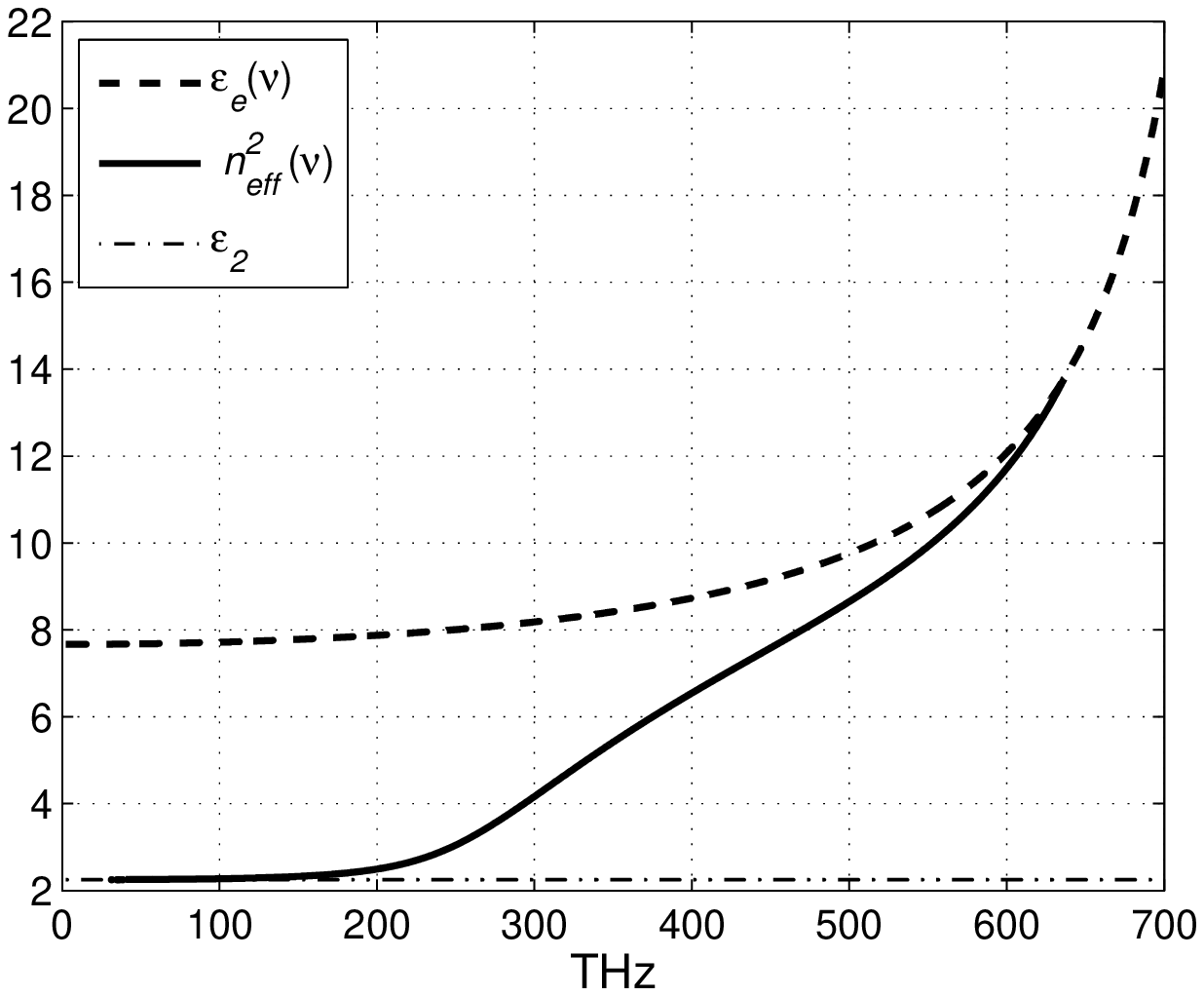}
        \\(\textit{a}) $\kappa = 5$
    \end{minipage}
    \vfill
    \begin{minipage}[ht]{.9\linewidth}
        \centering
        \includegraphics[width=\linewidth]{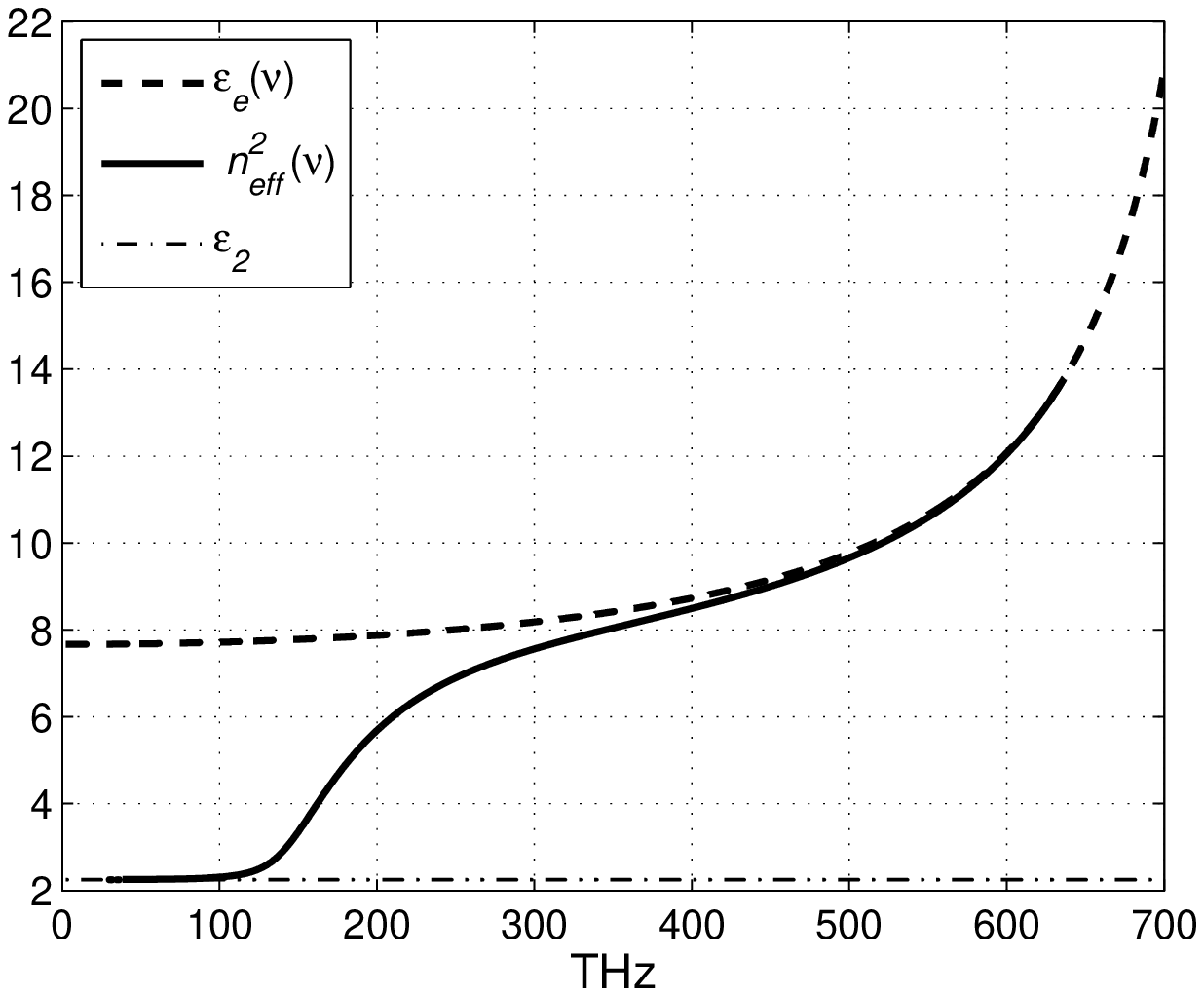}
        \\(\textit{b}) $\kappa = 10$
    \end{minipage}
    \vfill
    \begin{minipage}[ht]{.9\linewidth}
        \centering
        \includegraphics[width=\linewidth]{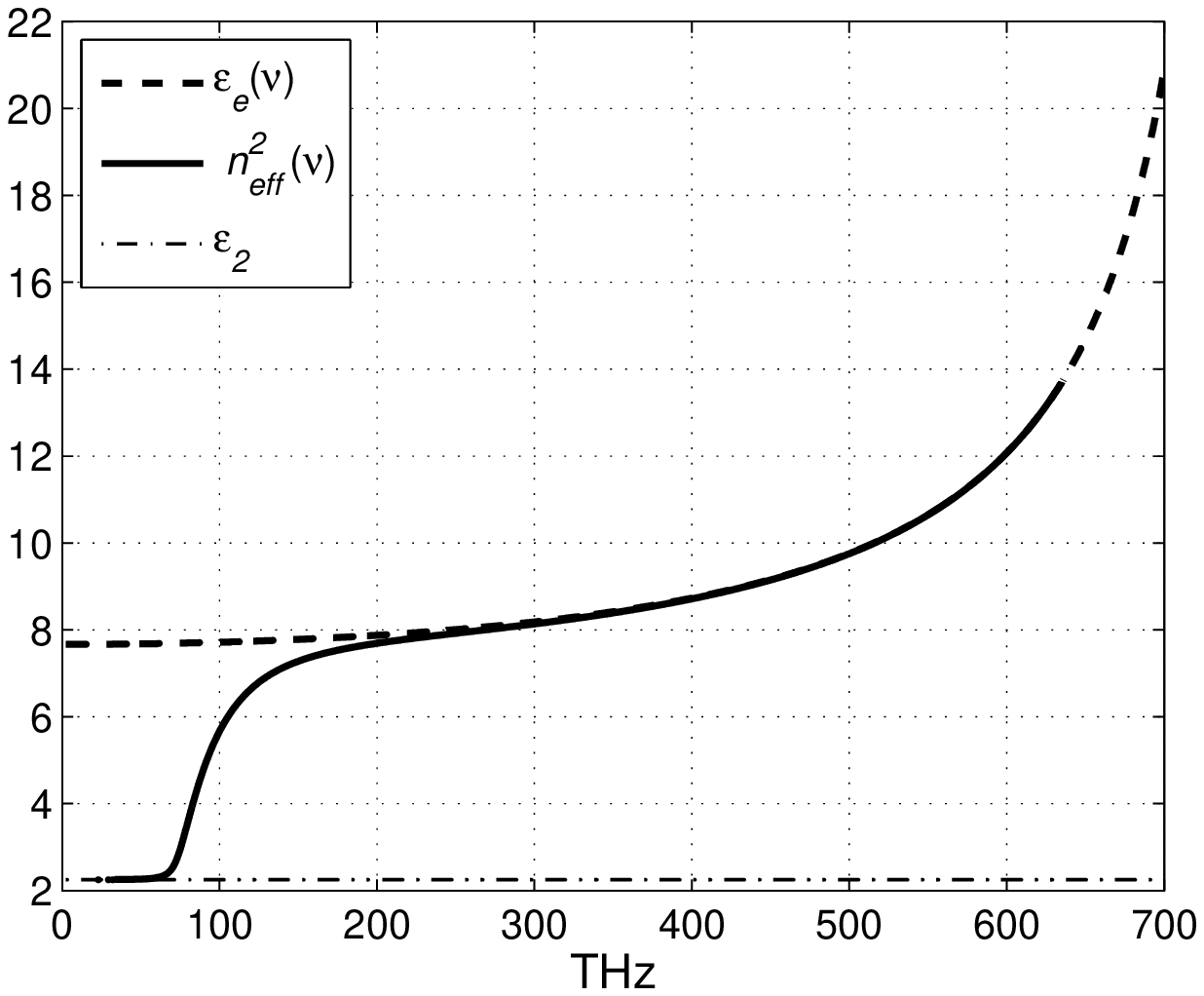}
        \\(\textit{c}) $\kappa = 20$
    \end{minipage}
    \caption{Dependence of $n_{eff}^2$ on frequency for interface with topologically
             nontrivial insulator, $\mathbf{l} = \mathbf{n}$.}
    \label{fig6}
\end{figure}

The frequency dispersion of hyperbolic medium permittivities
$\varepsilon_e(\omega)$, $\varepsilon_o(\omega)$ here is taken into
account with the same parameters as in subsection \textbf{A}.
Solutions of equation (\ref{eq:DispN}) for different values of
$\kappa$ are presented in Fig. \ref{fig5}, and \ref{fig6}.

As follows from figures $n_{eff}^2$ exists for whole frequency
range, where composite material described by (\ref{eq:EffectMedium})
is hyperbolic one with $\varepsilon_o(\omega) < 0$ and
$\varepsilon_e(\omega) > 0$ (compare with Fig. \ref{fig2}). One can
notice from Fig. (\ref{fig5}) ($a$), ($b$)  and Fig. \ref{fig6}
($a$) that point, where $n_{eff}^2 = \varepsilon_e(\omega)$, is the
same as that, were $\varepsilon_o(\omega)$ change its sign.

The considered situation is differ from the cases of surface waves
at topological insulator/ metal and dielectric/ metal interfaces.
The main feature is that frequency interval of surface wave
existence is limited before the critical frequency $\omega_c$ could
be achieved.

 The magnetoelectric constant $\kappa$ is taking equal
to $(2\times 100+1)~\alpha$. The big value of the axion field
$\theta$ is used to obtain better illustration of the dispersion
curves. Figures \ref{fig4}, \ref{fig5} and \ref{fig6} indicate that
the main features of the dispersion curves depend only slightly on
the $\kappa$. The magnetoelectric effect generate the TE-TM wave
coupling independently of the value $\kappa \neq 0$.

\section{The role of dissipation of the hyperbolic material}

\noindent The hyperbolic materials are composite materials that
contain conductive components such as metallic layers or nanoroads.
The metallic inclusions lead to the energy dissipation in these
materials. For real hyperbolic materials dielectric constants
$\varepsilon_o$ and $\varepsilon_e$ are complex values. For example,
 $\varepsilon_o = -2.78 +i 0.13$ and $\varepsilon_å =
6.31 + i0.09 $ at 465 nm \cite{Drachev:13}. The imaginary parts of
$\varepsilon_o$ and $\varepsilon_e$ can be small enough, if the
radiation frequency $\omega$ is far away from all typical for
material resonances.

The derivation of the dispersion relations in the case of complex
values of permittivities can be carried out as in the previous
sections. Metamaterial was described by permittivities
(\ref{eq:EffectMedium}). The contribution of the metallic inclusions
was taking into account according to Drude-Lorentz model with
$\gamma = 5.07 \cdot 10^{13} rad/c $.

The equations defining the dispersion relations have the forms that
are similar to (\ref{eq:DispTy}), (\ref{eq:DispTz}) and
(\ref{eq:DispN}). However, now the constants $\varepsilon_o$ and
$\varepsilon_e$ are complex values. Solutions of the obtained
equations $n_{eff}^2 (\omega)$ are complex ones.
$\mathrm{Re}~n_{eff}$ describes the velocity of the surface wave,
and $\mathrm{Im}~ n_{eff}$ describes the dumping this wave. The
reciprocal of dumping length is equal to $2(\omega/c)\mathrm{Im}~
n_{eff}$.

\begin{figure}[ht]
    \centering
    \includegraphics[width=.9\linewidth]{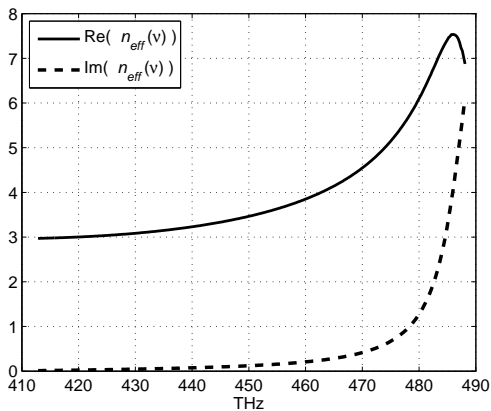}
    \caption{Dependence of $\mathrm{Re}~n_{eff}$ and $\mathrm{Im}~n_{eff}$ on frequency for
    the surface wave with regard to the dissipation.
    Optical axis is tangential to the interface. }
\label{LMG:Dissipa:1}
\end{figure}

\begin{figure}[ht]
    \centering
    \includegraphics[width=.9\linewidth]{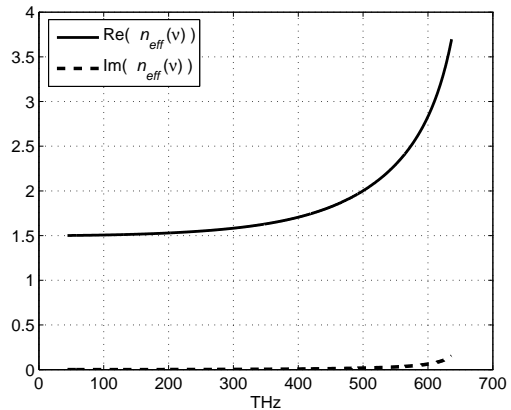}
    \caption{Dependence of $\mathrm{Re}~n_{eff}$ and $\mathrm{Im}~n_{eff}$ on frequency for
    the surface wave with regard to the dissipation.
     Optical axis is normal to the interface. }  \label{LMG:Dissipa:2}
\end{figure}

The Fig.\ref{LMG:Dissipa:1} and Fig.\ref{LMG:Dissipa:2} demonstrate
the dependencies of the real and imaginary parts of the effective
index in the different cases of the optical axis orientation. The
real parts of $n_{eff}^2$ in case of nonzero $\gamma$ have the same
shape as presented in figures \ref{fig4}, \ref{fig5} and \ref{fig6}.

The dissipative loss in the hyperbolic medium results in the finite
dumping length that is usual for plasmon-polariton surface waves.


\section{Conclusion}

\noindent Here the surface wave propagation along the interface
between a hyperbolic material and a topological insulator was
considered. The description of the electrodynamic of topological
insulators is based on the generalized Maxwell equations
\cite{Wilczek:87,XiaoLiangQi:08}. The continuity conditions for
tangent components of electric field vectors and the normal
components of induction vectors were used. The dispersion relations
for surface waves are derived. The cases of the anisotropy axes is
normal to interface or one is coplanar to interface are discussed.

If the optical axis of the hyperbolic material is tangential to the
interface and directed across the propagation direction, and the
magnetoelectric constant $\kappa$ is not zero, the surface wave
exists in the narrow frequency domain, if $\varepsilon_e >
\varepsilon_2$. In the case of trivial insulator ($\kappa=0$) the
surface wave corresponds to the conventional surface plasmon. It
should be note that the surface wave is impossible if the optical
axis aligned with propagation direction. This means that some
critical direction of the optical axis exists.

If the optical axis is normal to the interface the surface wave
exists for whole frequency range, where composite material is
hyperbolic one with $\varepsilon_o(\omega) < 0$ and
$\varepsilon_e(\omega) > 0$.

Due to mixing of TM and TE waves at topological insulator/hyperbolic
material interface the surface wave propagation conditions becomes
more strict comparing to topological insulator/metal or
dielectric/hyperbolic material cases. At last ones the surface wave
exists in the frequency interval $(0, \omega_c)$, at $\omega_c$
propagation constant tends to infinity. It is interesting to note,
that in case of topological insulator/hyperbolic material with
optical axis is tangential to the interface and directed across the
propagation direction this interval narrows to $(\omega_1,
\omega_c)$ if $\varepsilon_e > \varepsilon_2$. At $\omega_1$ the
propagation constant is $\beta(\omega_1) = k_0
\sqrt{\varepsilon_e}$. A value of $\omega_c$ is a function of
magnetoelectric constant. In case of the normal direction to
interface of the optical axis interval $(0, \omega_c)$ narrows to
$(0, \omega_2)$. At $\omega_2$ value of propagation constant is
finite $\beta = k_0 \sqrt{\varepsilon_e}$.

Here the topological insulator is considered. This material is
characterized by the axion field, which is static field. However,
axion field is static in a time-reversal invariant topological
insulator. In \cite{Li:Wang:Qi:10} the antiferromagnetic long-range
order in a topological insulator was discussed. As the result of
time-reversal symmetry breaking $\theta$ becomes a dynamical axion
field taking continuous values from $0$ to $2\pi$. As the dynamic
axion field couples nonlinearly to the electromagnetic field, this
term in the generalized Maxwell equations is the origin of the
nonlinear responses of the topological magnetic insulators. If there
is an externally applied static and uniform magnetic field, parallel
to the electric of the electromagnetic wave, the linear
approximation for the generalized Maxwell equations is
constructible. New type of the bulk polariton referred to as the
axionic polariton was proposed \cite{Li:Wang:Qi:10}. The surface
axionic polariton would be expected to be propagating along the
interface between a topological magnetic insulator and a
non-topological material.

\section*{ Acknowledgement}

This investigation is funded by the Russian Foundation for Basic
Research (Grant No. 15-02-02764).



\end{document}